# First Light Results from the Hermes Spectrograph at the AAT

Sheinis, Andrew [a, d]; Anguiano, Borja [g]; Asplund, Martin [b]; Bacigalupo, Carlos [g]; Barden, Sam [c]; Birchall, Michael [a]; Bland-Hawthorn, Joss [d]; Brzeski, Jurek [a]; Cannon, Russell [a]; Carollo, Daniela [g]; Case, Scott [a]; Casey, Andrew [j]; Churilov, Vladimir [a]; Couch Warrick [a]; Dean, Robert [a]; De Silva, Gayandhi [a]; D'Orazi, Valentina [g]; Duong, Ly [b]; Farrell, Tony [a]; Fiegert, Kristin [a]; Freeman, Kenneth [b]; Frost Gabriella [a]; Gers, Luke [a]; Goodwin, Michael [a]; Gray, Doug [a]; Green, Andrew [a]; Heald, Ron [a]; Heijmans, Jeroen [h]; Ireland, Michael [b]; Jones, Damien [e]; Kafle, Prajwal [l]; Keller, Stefan [b]; Klauser, Urs; Kondrat, Yuriy [a]; Kos, Janez [k]; Lawrence, Jon [a, g]; Lee, Steve [a]; Mali, Slavko [a]; Martell, Sarah [i]; Mathews, Darren [a]; Mayfield, Don [a]; Miziarski, Stan [a]; Muller, Rolf [a]; Pai, Naveen [a]; Patterson, Robert [a]; Penny, Ed [a]; Orr, David [a]; Schlesinger, Katharine [b]; Sharma, Sanjib [d]; Shortridge, Keith [a]; Simpson, Jeffrey [g]; Smedley, Scott [a]; Smith, Greg [a]; Stafford, Darren [a]; Staszak, Nicholas [a]; Vuong, Minh [a]; Waller, Lewis [a]; Wylie de Boer, Elizabeth [b]; Xavier, Pascal [a]; Zheng, Jessica [a]; Zhelem, Ross [a]; Zucker, Daniel [g,] Zwitter, Thomas [k]

[a] Australian Astronomical Observatory (AAO) PO Box 296, Epping NSW 1710, Australia
[b] Australian National University (ANU), Research School of Astronomy & Astrophysics, Mount Stromlo Observatory, Weston ACT 2611, Australia
[c] Leibniz-Institut für Astrophysik Potsdam (AIP),14482 Potsdam, Germany
[d] School of Physics, The University of Sydney NSW 2006, Australia
[e] Prime Optics, Eumundi, QLD 4562, Australia
[f] Centre for Astrophysics Research (CAR), School of Physics, Astronomy and Mathematics, University of Hertfordshire (Herts), Hatfield, AL10 9AB, UK
[g] Department of Physics and Astronomy, Macquarie University, NSW 2109, Australia
[h] TNO – Hoofddorp Netherlands
[i] School of Physics, University of New South Wales, Sydney NSW 2052, Australia
[j] Institute of Astronomy, University of Cambridge, Cambridge CB3 0HA, UK
[k] Faculty of Mathematics and Physics, University of Ljubljana, Slovenia
[l] International Centre for Radio Astronomy Research (ICRAR), The University of Western Australia

## ABSTRACT

The High Efficiency and Resolution Multi Element Spectrograph, HERMES is an facility-class optical spectrograph for the Anglo-Australian Telescope. It is designed primarily for Galactic Archaeology [21], the first major attempt to create a detailed understanding of galaxy formation and evolution by studying the history of our own galaxy, the Milky Way. The goal of the GALAH survey is to reconstruct the mass assembly history of the of the Milky Way, through a detailed chemical abundance study of one million stars. The spectrograph is based at the AAT and is fed by the existing 2dF robotic fiber positioning system. The spectrograph uses VPH gratings to achieve a spectral resolving power of 28,000 in standard mode and also provides a high-resolution mode ranging between 40,000 to 50,000 using a slit mask. The GALAH survey requires a SNR greater than 100 for a star brightness of V=14 in an exposure time of one hour. The total spectral coverage of the four channels is about 100nm between 370 and 1000nm for up to 392 simultaneous targets within the 2 degree field of view. HERMES has been commissioned over 3 runs, during bright time in October, November and December 2013, in parallel with the beginning of the GALAH Pilot survey, which started in November 2013. In this paper we present the first-light results from the commissioning run and the beginning of the GALAH Survey, including performance results such as throughput and resolution, as well as instrument reliability.

**Keywords:** HERMES, spectrograph, AAT, 2dF, VPH, fiber

*asheinis@aao.gov.au; phone +61 2 93724821; fax +61 2 93724880; www.aao.gov.au

## 1. INTRODUCTION

The latest in a long history of instruments to be developed by the AAO Instrumentation Group (IG) for the Anglo-Australian Telescope (AAT) is the High Efficiency and Resolution Multi-Object Spectrograph (HERMES) for the AAT. HERMES is a facility instrument, designed to provide high-resolution multi-object spectra in the visible. The primary science driver is Galactic Archaeology, in which detailed abundances of up to 29 elements are used to chemically tag stars in order to understand their origin and star formation history. The primary program is the Galactic Archaeology with HERMES survey (GALAH), which has begun as a pilot survey in late 2013 and is now in full data gathering operations. GALAH will measure the chemical abundances of 1,000,000 stars in the Milky Way. HERMES provides a nominal spectral resolving power of 28,000 for GALAH with a high-resolution mode of about 45,000, over 4 non-contiguous bands within the 370-1000 nm window. The opto-mechanical design of HERMES[2] allows for reconfiguration of the bands to enable astronomers to pursue other spectral bands of interest between 370 - 1000 nm. The spectrograph uses four large 500 x 200mm Volume Phase Holographic (VPH) gratings and is fed by the 2dF fiber-positioning robot at the AAT telescope prime focus. There are two slits of 392 science fibers each that allow for one slit to collect science data while the robot configures the fibres feeding the other slit for the next observation. Details of the 2dF positioner system are available in [23].

The AAO IG has: completed its assembly, integration and testing of the HERMES spectrograph at the AAO head quarters in Sydney; delivered the instrument to the AAT site in mid 2013; and along with the GALAH team, has fully commissioned the instrument in Oct-Dec 2013.

## 2. TOP LEVEL SPECIFICATIONS

The HERMES spectrograph is required to achieve the following top-level functional performance specifications:

1) The HERMES system should have high spectral resolution ($\lambda/\Delta\lambda$) of ~50,000 at multiplex of 400 targets. (For Stellar Astrophysics, Interstellar Medium, and Radial Velocity Variability science cases.)
2) The HERMES system shall provide 4 configurable windows located in the wavelength range 370 - 1000 nm. Nominally:
   a. 471.8 nm – 490.3 nm
   b. 564.9 nm – 587.3 nm
   c. 648.1 nm – 673.9 nm
   d. 759.0 nm – 789.0 nm
3) The HERMES system shall provide the following sensitivities for brightness: V=14, SNR of 100 in 60 minutes of integration. For Galactic Archaeology (V=14) this corresponds to a system efficiency of 0.1
4) Brightness of targets: 10–14 for main Galactic Archaeology survey; down to 16 – 17 for targeted observation.

## 3. ASSEMBLY AND INTEGRATION

At the end of March 2013, initial integration and testing, in the Epping-Sydney lab, of the HERMES Spectrograph came to a close. Tests were carried out on a fully aligned spectrograph with completed Blue, Green, and Red Channels. The IR Channel, while structurally complete, was not tested as the IR-VPH grating was still being manufactured. Testing in the Epping lab was limited to the use of a test fixture for the slit, which is similar to, but does not perform to the full standard of the slits attached to the 2dF Fiber Cable. Included in the testing was the acquisition and reduction of a solar spectrum by illuminating a test fiber bundle with sunlight and feeding that to HERMES. The testing showed conclusively that all the systems within HERMES were operational and ready for shipment to the AAT at Siding Spring Observatory, but that the optical (image-quality) performance in the green and red channels was not within spec. After extensive testing it was determined that this was primarily due to wavefront errors (primarily astigmatism) introduced in the green and red gratings. It was decided at the time that this issue should not drive the overall HERMES Schedule and the issue could be resolved in parallel. In addition, the AAO instrumentation group was required to vacate our Epping facility to move and re-establish ourselves in our new facility in North Ryde, some 5 Km from Epping. Our time in Epping had come to a close.

In the first week of April 2013 the process of disassembling HERMES, packing, and transporting to site began. During the disassembly process HERMES received its NEXTEL black interior paint work. After 5 weeks, five 20-foot container truck loads, 6 moving van loads, and approximately 10 station wagon trips, HERMES was safely transported to the AAT.

On the last week of June 2013, the re-assembly process began at the AAT in Coonabarabran NSW. The aggressive goal was to have a fully operational instrument for the scheduled first commissioning run on October 19$^{th}$, 2013. This left little over 3 ½ months to take a completely disassembled and packaged HERMES to a fully rebuilt, aligned, working instrument. This was quite a challenge but the team was keen.

After a complete design, and fabrication stage at site, the HERMES temperature-controlled room was ready to receive its instrument. The first step was to position the vibration isolators in the correct position on top of the concrete piers. The next step of the rebuild process was to re-assemble the HERMES structure. This main support structure is called the "spine" of HERMES. The spine supports the slit assembly, field lens, collimator mirror, corrector lenses, and beam splitters. There are four structures that are bolted and pinned to the spine, one for each arm of the spectrograph. These structures support a grating, fold mirror, and light baffling. Cantilevered off each of the four channel structures are the camera focus drive assemblies and cryostat assemblies. The HERMES structure was designed to be modular to aid in the disassembly and re-assembly process.

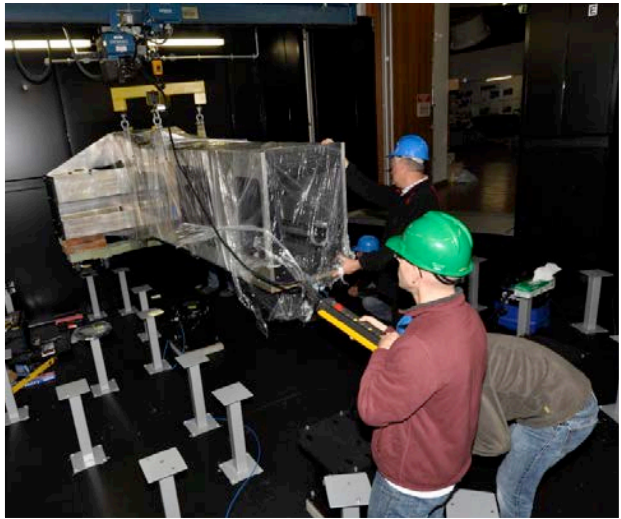 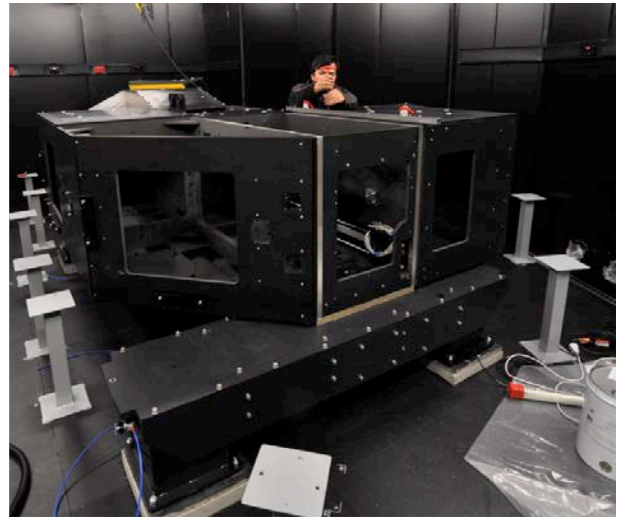

*Figure 1. HERMES Spine being set to air isolators.*
*Figure 2. HERMES structure nearing final assembly. IR Box removed to allow collimator alignment.*

Following rebuild of the HERMES structure, the blue, green, and red camera assemblies were re-attached. To save time and schedule the cameras were shipped fully integrated in their structural mount/flexural focus drive mechanism. With the cameras mounted, the 1 meter-diameter HERMES collimator mirror was reassembled into its mount and then reset onto the structure. All the most massive HERMES components were now set to the frame, fully loading the structure, with the exception of the IR channel structure and camera, which was left until later in the assembly to allow alignment access to the collimator down the center of the spine

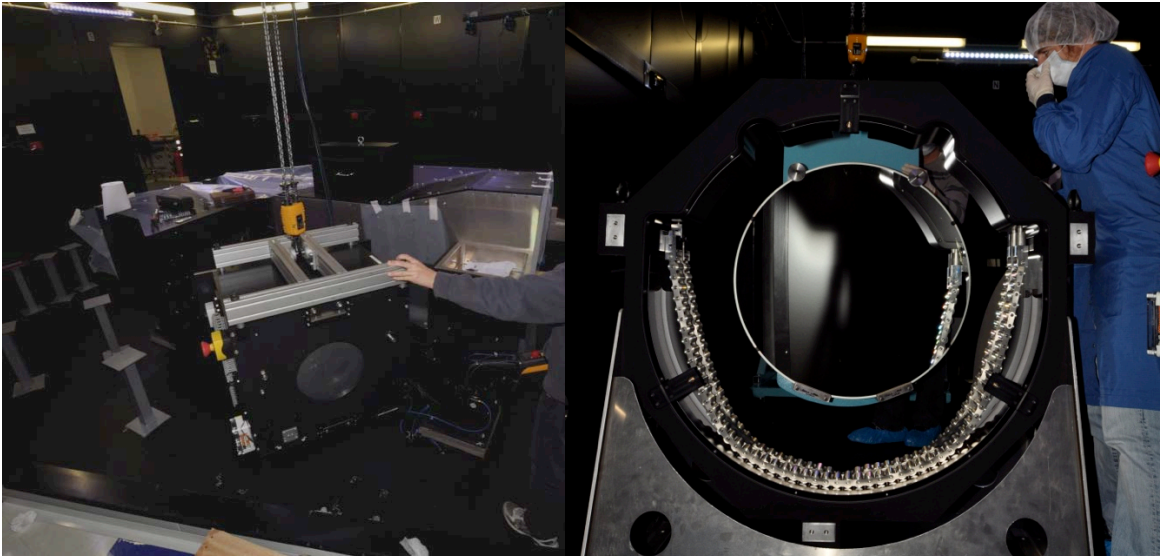

*Figure 3. Green camera assembly being craned into position.*
*Figure 4. Collimator mirror being inserted into its chain mount.*

The collimator alignment process was now able to commence with the HERMES collimator mirror first being set to the established optical axis[20]. The collimator mirror was tested interferometrically in the assembled condition to ensure no stress was being transferred to the surface and it was still meeting wavefront requirement. Following the collimator mirror alignment the field lens and corrector lens were aligned to the optical axis. The slit assembly was set into position and focus was verified across all fibers of the two slit assemblies, as we looked back through the collimator assembly with a metrology telescope arangement.

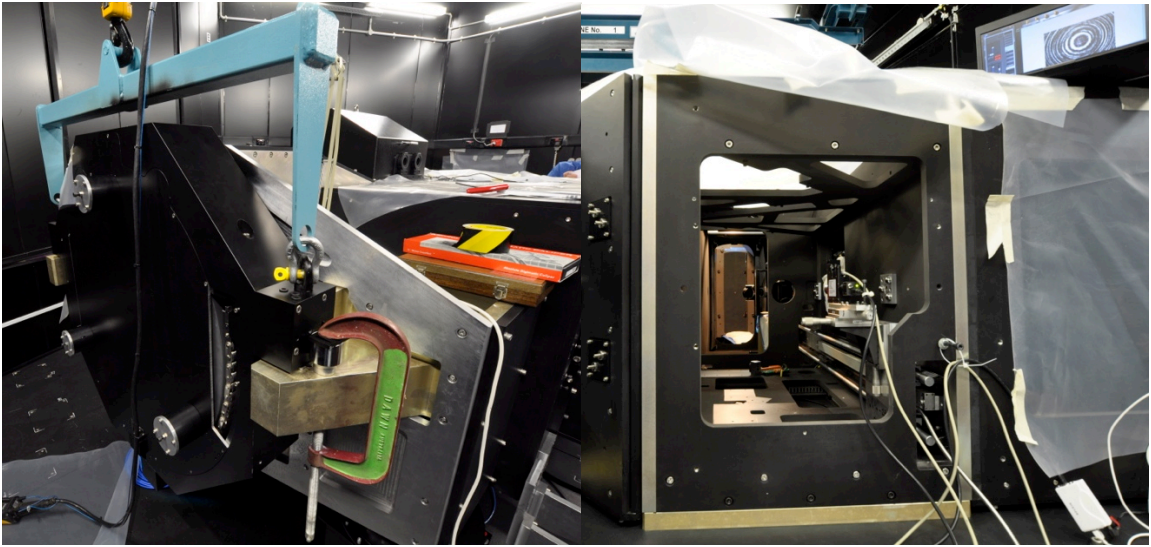

*Figure 5. Collimator mirror craned on top of the spine mount.*
*Figure 6. View through the spine. Point source microscope in position to align collimator optics.*

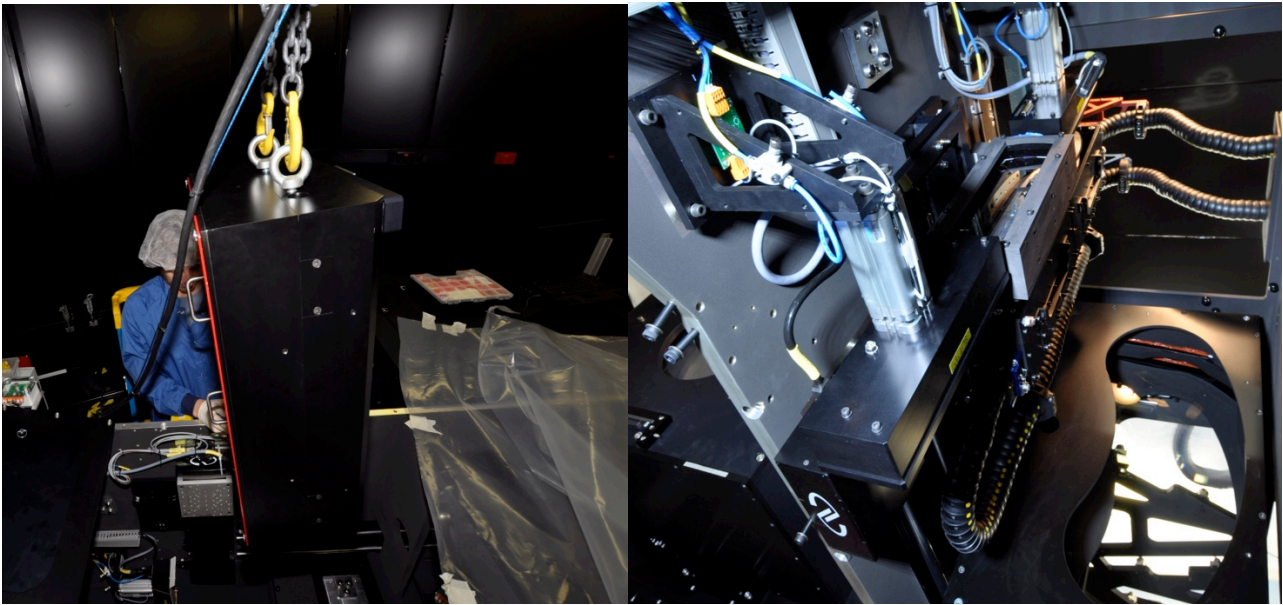

*Figure 7. Corrector lens being lowered into their kinematic mounts within the frame.*
*Figure 8. View of the exchanging slit assembly, field lens, back illuminators, and corrector lens asembly (right).*

In parallel with the HERMES re-construction was an effort to solve the grating astigmatism issue. As detailed in section 4.3, the plan was to repolish the gratings flat and have them re-coated. This plan went forward and sucessfully brought the gratings to a completed state just as they were needed for the HERMES channel alignment.

With the HERMES collimator assembly fully assembled and aligned attention was turned to aligning each of the fold paths. Alignment progressed from Blue, Green, Red and finally the IR path. Using prism setups, alignment telescope and PSM, each of the fold paths were brought into alignment with the previously established optical axis[20]. The beam splitters and gratings were adjusted via shimming on their kinematic mounting arrangement. Fold mirrors utilized an integrated precision tip-tilt arangment in the mount driven by high precision 100 pitch screws. Alignment targets to each of the cameras were previously set, and camera alignment to the optical axis was verified to be within specification (1 arcminute).

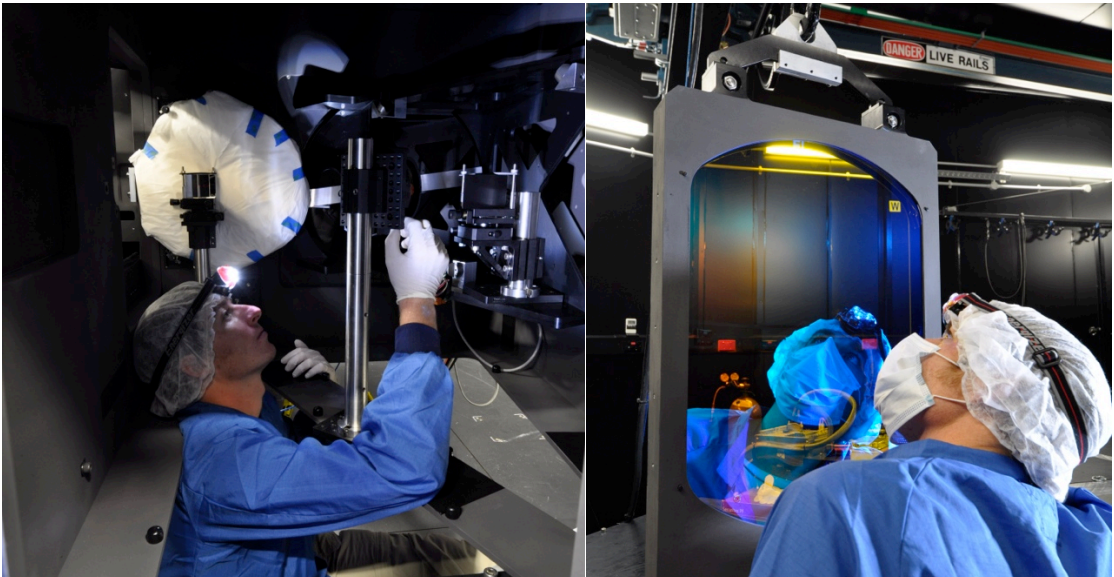

*Figure 9. Green channel alignment setup.*

*Figure 10. Beamsplitter being craned into position.*

In parallel with the opto-mechanical assembly and alignment the electronics team re-assembled all the electronic control cabinets, ran all the cables to the instrument, and fully re-wired HERMES. Cryostats were remounted after channel alignment. Controllers were assembled to the cryostats. The process of pumping each cryostat and cooling to temperature began. After each cryostat was brought to temperature final alignment of each cryostat to each camera was performed. Cameras were focused utilizing every last minute, but HERMES was ready and saw first light on the well-studied globular cluster 47 Tuc on its scheduled commissioing run on Ocotber 19$^{th}$ *having been finished with just hours to spare*!

Since the first commissioning run, HERMES has completed the GALAH pilot survey, and is now taking scientific survey data. The HERMES room is now assembled and complete. Only minor tweaks to the alignment have been implemented. Minor bug fixes have been implemented. The instrument is now fully in the hands of the AAT site staff.

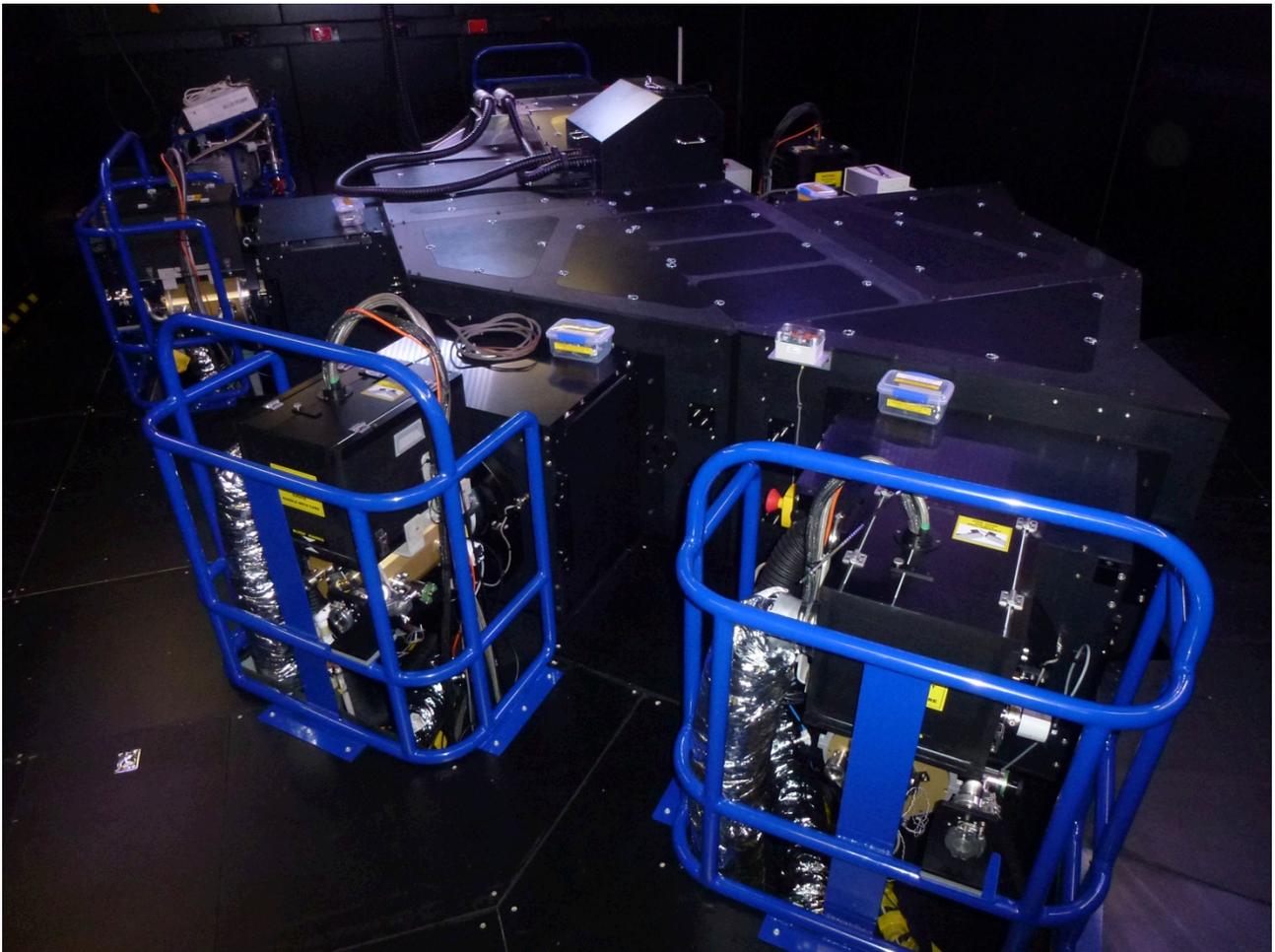

*Figure 11. HERMES fully assembled.*

## 4. OPTICAL COMPONENTS

There are three main optical assemblies in the HERMES spectrograph that were deemed high risk. While in a spectrograph as large as HERMES all optics are high risk the camera assemblies, collimator assembly, and the especially the volume phase holographic grating (VPH) had the most potential to affect performance and delivery schedule. As a result a series of metrology tests were done to ensure that vendors delivered as promised and to minimize for any unaccounted performance losses in the spectrograph.

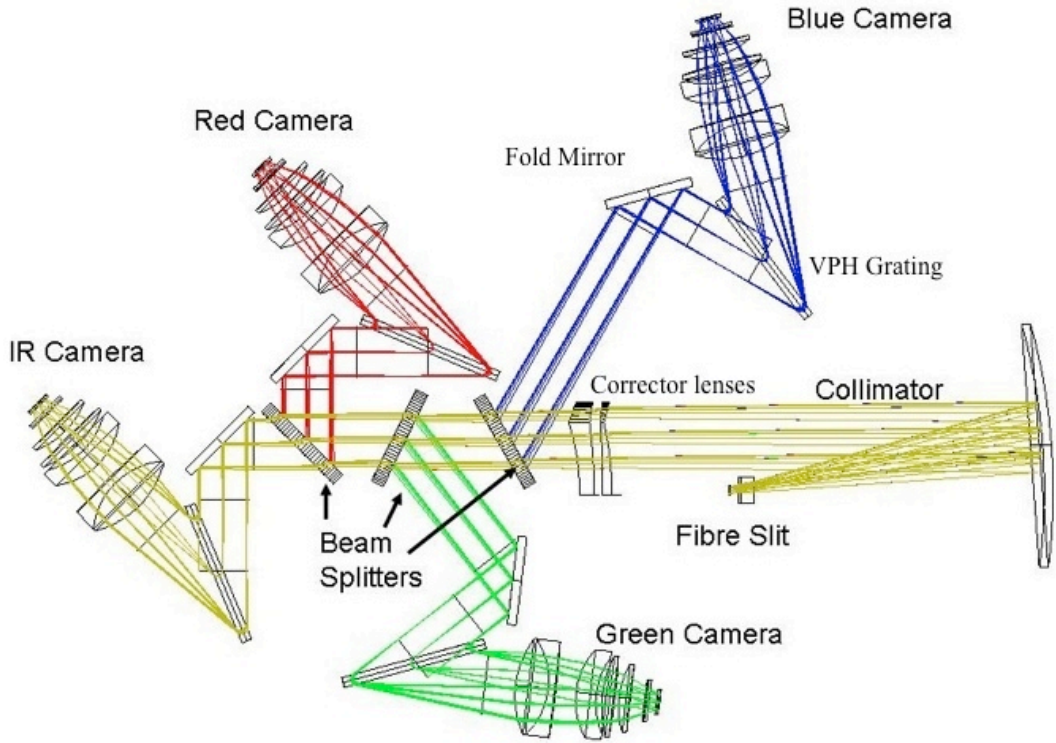

*Figure 12. Optical layout of the HERMES spectrograph*

**4.1 Camera optics**

Each of the four cameras in the spectrograph is an F/1.67 refractive design with a 380 mm diameter first element, with an entrance pupil of 190 mm located at the grating. The camera assembly contains a doublet, with an aspheric first surface, and then 3 additional lens elements. The camera assembly does not contain the dewar optics, which consist of 2 windows and a field flattener lens just above the detector. The vendor supplying the cameras relied on a coordinate measuring machine data (CMM) to align the individual lenses of the camera in the lens cell, which was determined to be insufficient to fully verify performance. To ensure the performance of the cameras two tests were conducted by AAO personnel on the cameras: a double pass interferometric test and a lens alignment check with an autocollimating alignment telescope at the vendor facility and again after shipping to the AAO.

Each camera was designed to overlap the previous camera's wavelength coverage for future instrument modifications. While this made the design a little more challenging this did allow a common null lens to be used for the interferometric testing with only slight modifications to the distance between the null lens, the camera, and the interferometer. Due to the weight of the camera, 220 kg, the interferometric testing was designed in such a way that the interferometer and the null lens were brought to the optical axis of the camera. The interferometer was attached to a 250 mm precision linear stage with an F/1.5 reference sphere. The interferometer was aligned iteratively using the returns from camera lens optical surfaces at their respective centers of curvature (COC). By setting the first surface of the of the fourth lens as a reference as well as the last surface on the third the lens, an optical axis was quickly established with minimal distortion from other elements in the system. Furthermore, the inteferometric returns confirmed the lens spacings and defined offset positions for the null lens. Using a Monte-Carlo tolerance analysis the upper and lower bounds were predetermined for a number of Zernike polynomials that would provide a Go/No-Go performance evaluation for the cameras, as is typically done for mass produced components. If the inteferometric test proved the performance insufficient then the alignment of each lens could be determined using the autocollimating alignment telescope with the PIP attachment[20]. The second test uses the autocollimating alignment telescope with a fixed point source set at a distance from the optical assembly. The high intensity point source is reimaged by each optical surface. By aligning the alignment telescope to two optical surfaces in the camera the displacement of all the surfaces could be determined using the 'pip' returns from surface whether real or virtual. Fortunately, the cameras tested well the first time at the vendor facility and again at the AAO requiring only a quick check using the second test method.

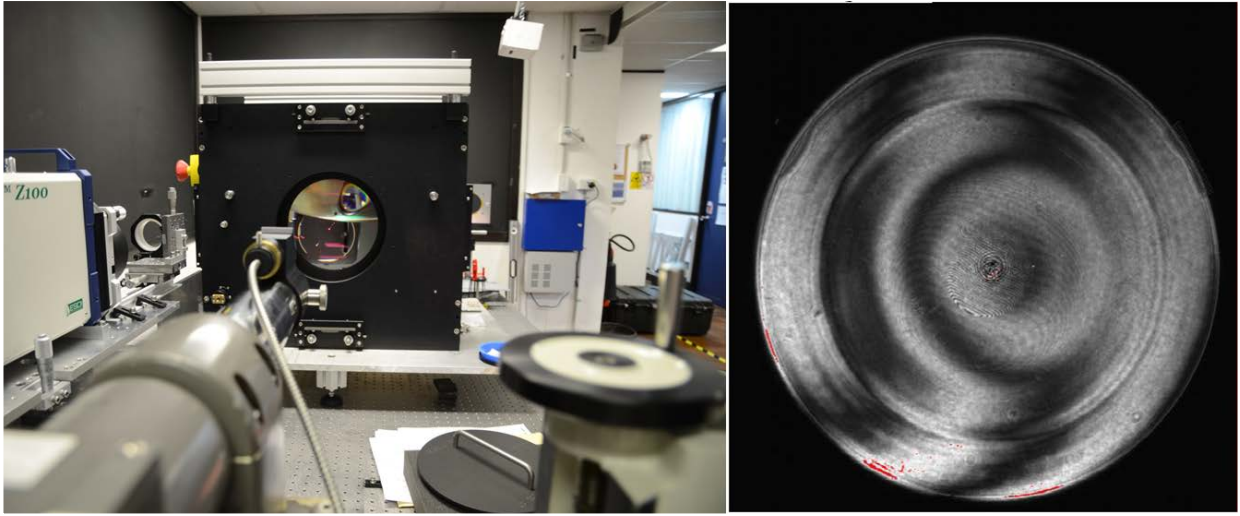

*Figure 13. Camera under test with the autocollimating alignment telescope after inferometric testing (left), IR camera transmitted wavefront using the null lens resulting in a measured wavefront of RMS 0.17 wv and PV 1.88 wv at 632.8 nm (right).*

**4.2 Collimator assembly**

The HERMES spectrograph uses an off-axis Houghton-based collimator design with a slit curved to match the radius of the focal surface as presented to the spectrograph and curved again perpendicularly along the surface of the sphere to compensate for the spectral smile. The collimator consists of a relatively small field lens, a 1 meter diameter spherical mirror weighing 500 kg, and two off-axis corrector lenses 620 x 330 mm in diameter in an assembly weighing 250 kg. The collimator optics each had their issues. While each issue was found in testing, some were expected due to good reporting from vendors, and others were uncovered later in the process.

The 1 meter diameter collimator mirror is a standard 6:1 diameter to thickness ratio mirror with enhanced silver coatings to improve throughput at 370 nm. The mirror had two issues. One, the mirror is strap mounted using an invar and stainless steel chain. This made the mirror very sensitive to gravitational loading, created if the mirror was unbalanced with respect to the gravity vector and the face contact pads. This required interferometric testing at the time of integration to qualify the angle of the mirror with respect to gravity so the mounting didn't inadvertently load the mirror face on the front reference pads. Two, the enhanced silver coating has proved challenging on two fronts. The coating performance proved very challenging to obtain on such a large mirror and the adherence of the coating is proving softer than promised. While witness samples of the coating passed testing for adhesion and abrasion requirements using MIL-C-48497, the coating on the mirror itself has proven soft and uncleanable as it does not conform to MIL-C-48497.

The fabrication and alignment of the collimator corrector lenses proved difficult from the onset. Due to the sheer size of the refractive off-axis element, it was deemed not cost effective and technologically too challenging to make a full-size on-axis parent. Thus, the blank size was restricted to the final as-built size of the lens to reduce costs. This made the fabrication of the second surface on each corrector lens particularly challenging as the surface became a tilted/decentered spherical surface with respect the first surface. The alignment by the vendor relied on a large three axis CMM to square each surface in assembly. It was first thought that this would provide an accurate alignment even with the limited number of sampling points at each step in the alignment of the two lenses in a common cell. The long radius of curvature on two of the surfaces, 3813 mm and 1986 mm, the sagittal radii difference later proved to be too small to discern over the error of the machine for a given sampling area, the sampling area too small, and possibly that a mounting flange ground into one of the lenses had the wrong angle. This resulted in an alignment error of the two lenses with respect to each other far above the required tolerance. The error was found using the autocollimating alignment telescope and the PIP attachment to image the centers of curvature of each lens surface. Further analysis in optical design software showed that the error added little change to the performance of the system. Due to time constraints it was deemed a worthy exercise to integrate the corrector lenses into the system as long as the camera optics, collimator mirror, and the gratings induced minimal error. With the errors known for the corrector assembly, the alignment compensated for the errors in order to locate and align the collimator corrector lenses. .

### 4.3 VPH grating

The VPHG proved not only a challenge to manufacture, but also to qualify the error induced by manufacturing errors. There are four gratings in HERMES, each 550 x 220 x 40 mm in size, with line frequencies from 3827 to 2378 lines/mm working at a 67.2 degree angle of incidence.

As highlighted in earlier papers[1,2,3,4,18,19], the extreme working angle of the grating forced the design of the grating to be optimized for s-polarization. Due to the size, the working angle of the grating, and the high resolution a custom large-aperture metrology system had to be built at the AAO to test the gratings for spectral and wavefront performance. Due to manufacturing processes the deformation of the carrier glass is very hard to control due to the internal stresses from the gelatin and bonding process. As a result the HERMES gratings had to be post polished after manufacturing to remove the deformation from the outer surfaces. What is still not clear is whether the grating surface inside the VPHG is itself bent along the internal surface and how that is contributing to wavefront error in the system due to our limited interferometer aperture. Due to time and budgetary constraints the gratings were integrated into the spectrograph and fortuitously, the maximum amplitude of diffracted wavefront error was lower than extrapolated error from sub-aperture measurements. Making the carrier glass thicker is one way to solve the problem, but due to the size of the grating in HERMES the glass weight quickly becomes unmanageable and would require complete retooling at the manufacturer facilities if the glass diameter to thickness is made on the standard 6:1 ratio with respect to the long axis of the diameter. Another unexpected surprise for future spectrographs of equivalent size to HERMES using large VPH gratings is the number of carrier plates needed during the manufacturing process. The AAO purchased 13 plates for only 4 gratings.

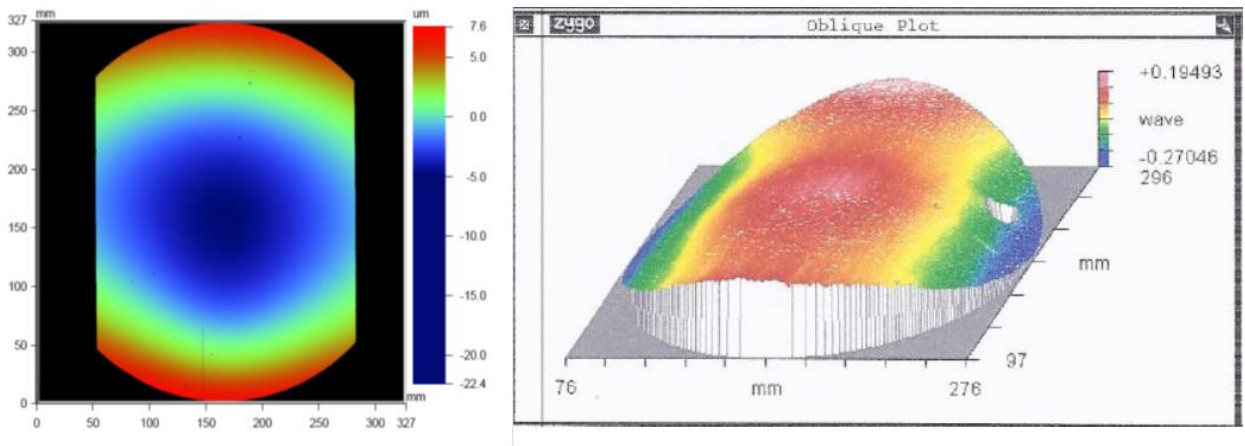

*Figure 14. Red VPHG surface profile before repolishing (left), Red VPHG transmitted wavefront at repolish at working angle of 62 degrees. (right).*

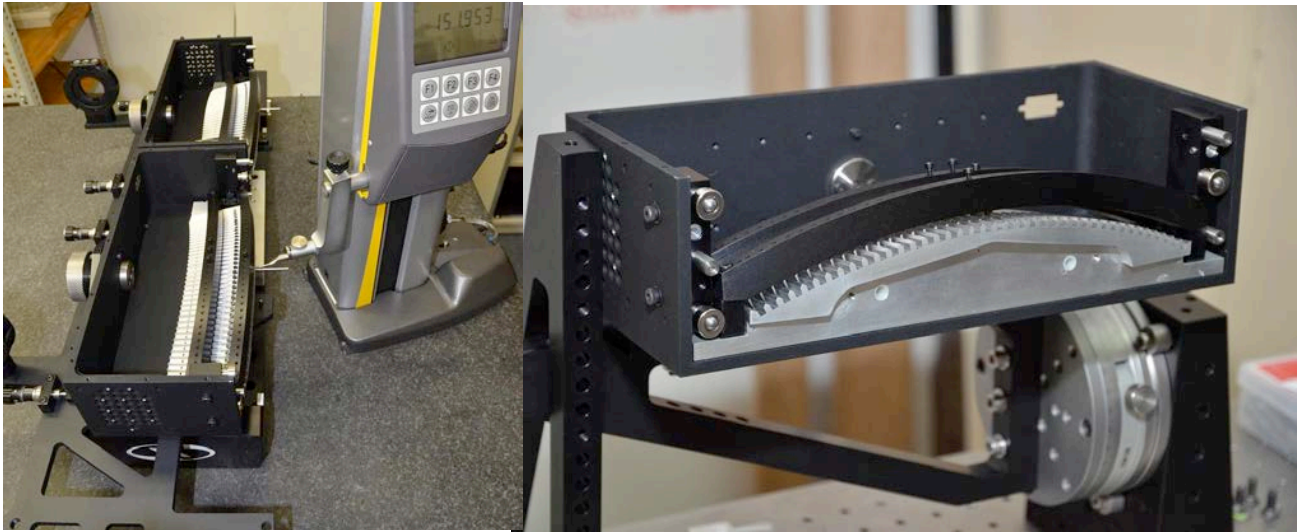

*Figure 15. The two curved slit assemblies on the translation stage, right: single V-groove block for lens mounting 40 lens sets*

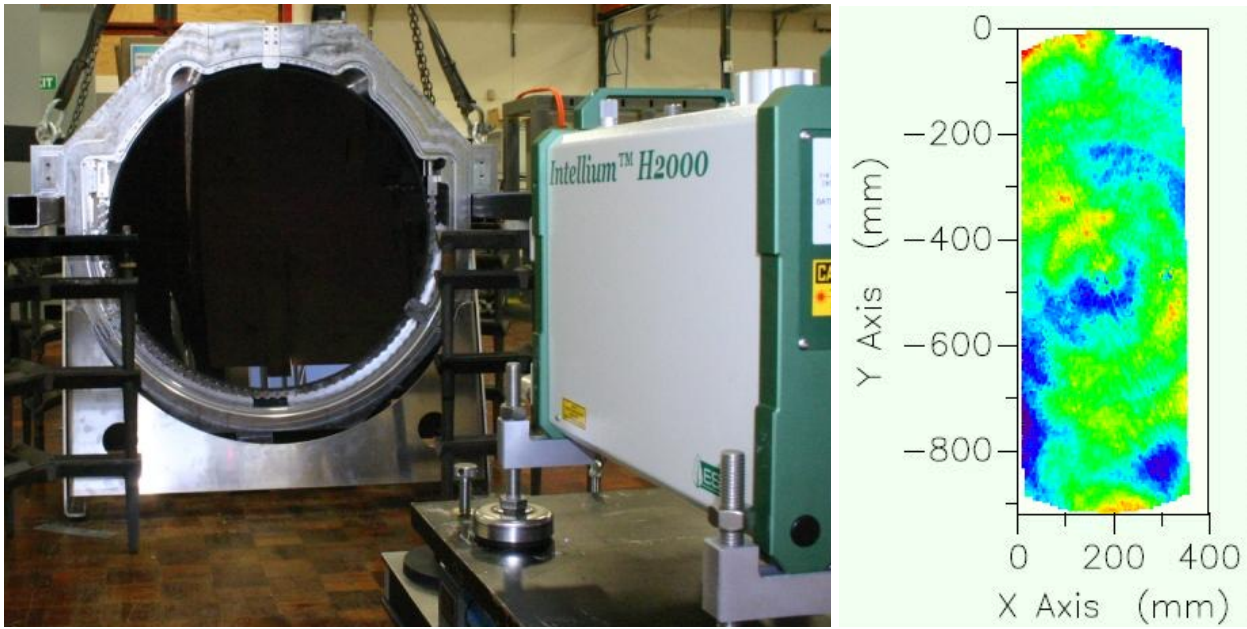

*Figure 16. The mounted collimator mirror before coating and its interferogram, P-V=0.1 wave, RMS=0.013wave*

# 5. CRYOSTATS AND DETECTORS

The HERMES detector cryostat shell is based on an Infrared Laboratories cryostat with a Polycold PCC cryocooler cold head to which we have fitted a Pfeiffer PKR full range vacuum gauge head. The end plate is replaced by a custom shell extension which encloses the detector and field flattener assembly. Hermetically sealed connectors, one for the detector control and data, another for thermal control, are mounted through the wall of this shell extension. To maintain thermal isolation a G10 fiberglass 'spider' is used to locate the detector and field flattener to the wall.

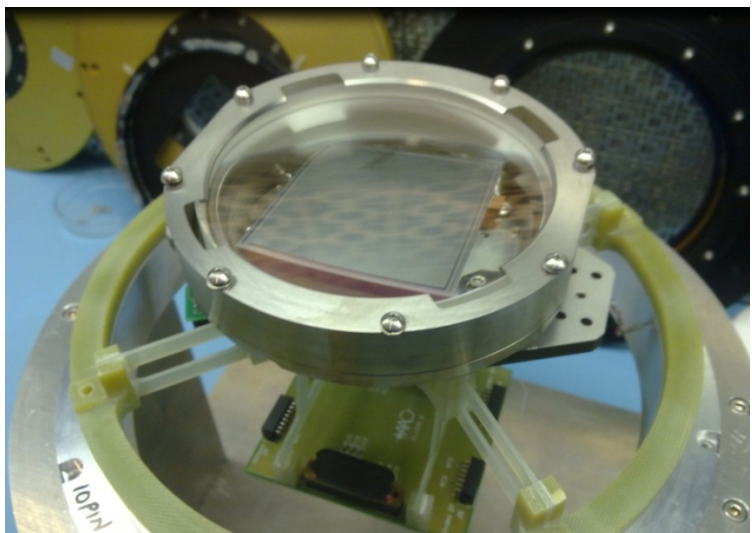

*Figure 17. The detector assembly with the mounted field lens mounted, (before blackening)*

To ensure a good match to the thermal expansion of the CCD detector, a molybdenum mounting plate is used. The thermal control hardware, heaters and a diode sensor are mounted directly on this plate. Three heaters are used to provide an even distribution of the heat input required for temperature trimming to much less than 0.1°C.

The mount for the bi-concave field flattener lens is mounted with less than 4mm gap from the detector. Cooling of the detector assembly is implemented by linking the detector mounting plate to the cryocooler head using oxygen-free copper thermal links. To minimize thermal strains and maintain detector alignment these links incorporate corrugated sections which structurally decouple the cryocooler head from the detector assembly. The contamination 'getter' is also connected via separate links to the cold head. A radiation shield encloses the detector and field flattener assembly and the thermal link and getter assembly. The cryostat shell is sealed by a front plate which accommodates the plane window, which withstands the atmospheric pressure load and a second thinner window. Dry nitrogen may be flushed between these windows to prevent dewing that might otherwise occur on high-humidity nights.

Internal electrical wiring is accomplished using flexible PCBs and the detector controller electronics box is mounted directly on the cryostat shell to minimize noise.

Design variations between cryostats result from the different wavelength ranges at which they operate and are limited to different CCDs, field flatteners and window coatings.

The detector control electronics includes the CCD controllers and their power supplies for each detector, as well as the computer systems that send commands to and acquire data from the CCD controllers. Each CCD detector in HERMES is controlled with an AAO2 CCD controller[5]. These controllers are configured for operation with the e2v CCD231-84 detectors to permit readout from one, two or four detector outputs, at various readout rates with windowing and binning.

# 6. FIBER CABLE FEED

The HERMES/AAOmega fiber bundle was designed to co-inhabit the 2dF robot positioner so that no instrument changeovers were required to switch between HERMES and AAOmega, and also that there is no need for connectorisation[6]. This led to the design of two fibers occupying one magnetic button, one for AAOmega fiber and one

HERMES fiber, as shown in Figure 18 below. The 2dF robot positions the magnetic button on the field plate and determines its position by looking at the back illuminated science fiber. .

The existing cable had a fringing problem that caused a wavelength-dependent intensity fluctuation in the measured spectra. This was found to be related with a cavity that was formed between the prism and the fiber face at the button end of the fiber cable. This connection has been redesigned and adhesives are used that have a better thermal match, low shrinkage and creates a strong bond. The assembly was tested on the AAT for five months and no fringing was observed.

Another point of attention has been to prevent the fibers from stressing as this can cause focal ratio degradation, (FRD). With a fiber cable of almost 50 meters in length there are many points where the fiber is bent and fixated. With large bending radii, protective sleeves and low stress points at fixations, the FRD was kept low and the achieved throughput for the new fiber cable is significantly higher (>50% more) than the previous cable.

The 2dF buttons are comprised of fibers terminated in a glass ferrule with a polyimide jacket covering the two AAOmega/HERMES science fibers plus a short (30mm long) dummy fiber to align the fibers perpendicular to the prism face. These ferrules are then glued into magnetic buttons to locate them on the 2dF field plate. The images below show the assembly.

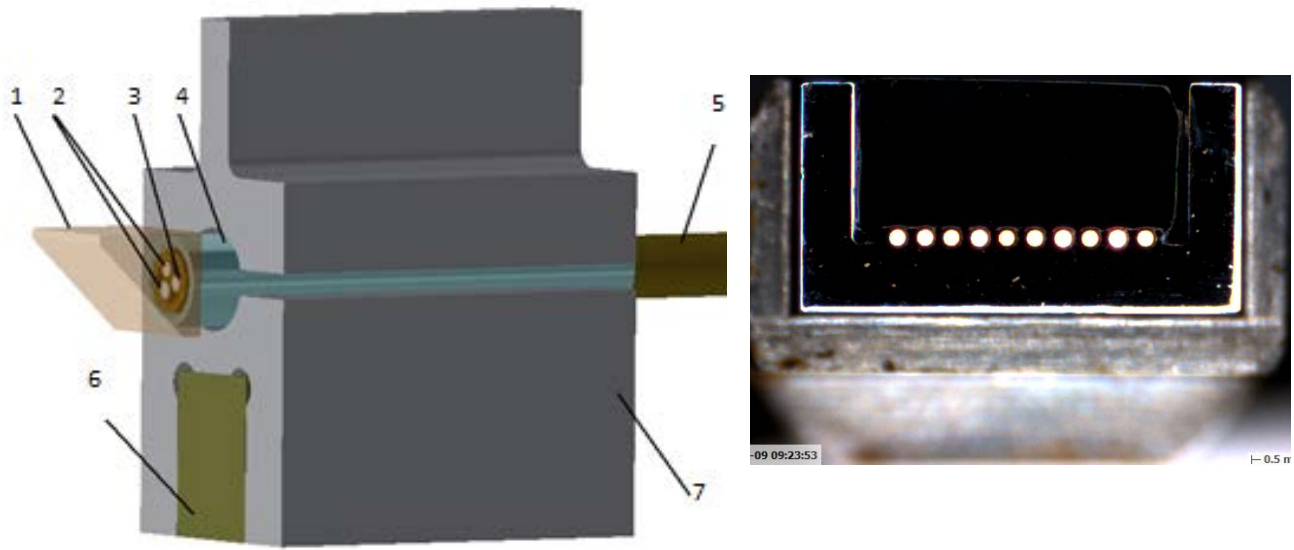

*Figure18. 2dF button. 1- Prism. 2- AAOmega/HERMES science fiber. 3- Dummy fiber. 4- Glass ferrule. 5- Polyimide tube. 6- Magnet. 7- Button body (Prism size is 1mm). Right: an image of 10 fibers in a slitlet*

At the slit entrance the fibers are assembled in slitlets that locate 10 fibers in v-grooves and collectively they make up a slit of 400 fibers.

The HERMES slitlets were glued and polished successfully. The locations of the fibers in their v-grooves with respect to their nominal positions relative to the slitlet body were measured using a MicroVu co-ordinate measurement system. The fiber locations were within the required radial position tolerance of +/-10μm to prevent crosstalk in the detector images.

## 7. CONTROLLERS AND ELECTRONICS

The electronics for HERMES is broadly separated into two categories: instrument control electronics and detector control electronics, as shown in Figure 10. The HERMES instrument control electronics is responsible for controlling and monitoring all the actuators and sensors used to configure and operate the instrument. The instrument control electronics is based on a distributed control system that uses the CANopen protocol running on the Controller Area Network (CAN) field bus. All analog and digital functions are interfaced to CANopen input/output nodes and all servo motor driven actuators are controlled by CANopen digital servo drives. The CAN bus is connected to an industrial control cabinet PC with a CAN interface, running the Linux operating system and operating as the CANopen master.

The instrument control electronics is implemented in six industrial electronics enclosures distributed around the outside of the spectrograph thermal enclosure. The majority of the instrument control electronics is commercial off-the-shelf. Some customization has been done on the slit back illumination control and drive, interlocking and interfacing for the Bonn shutters and servo amplifier breakout.

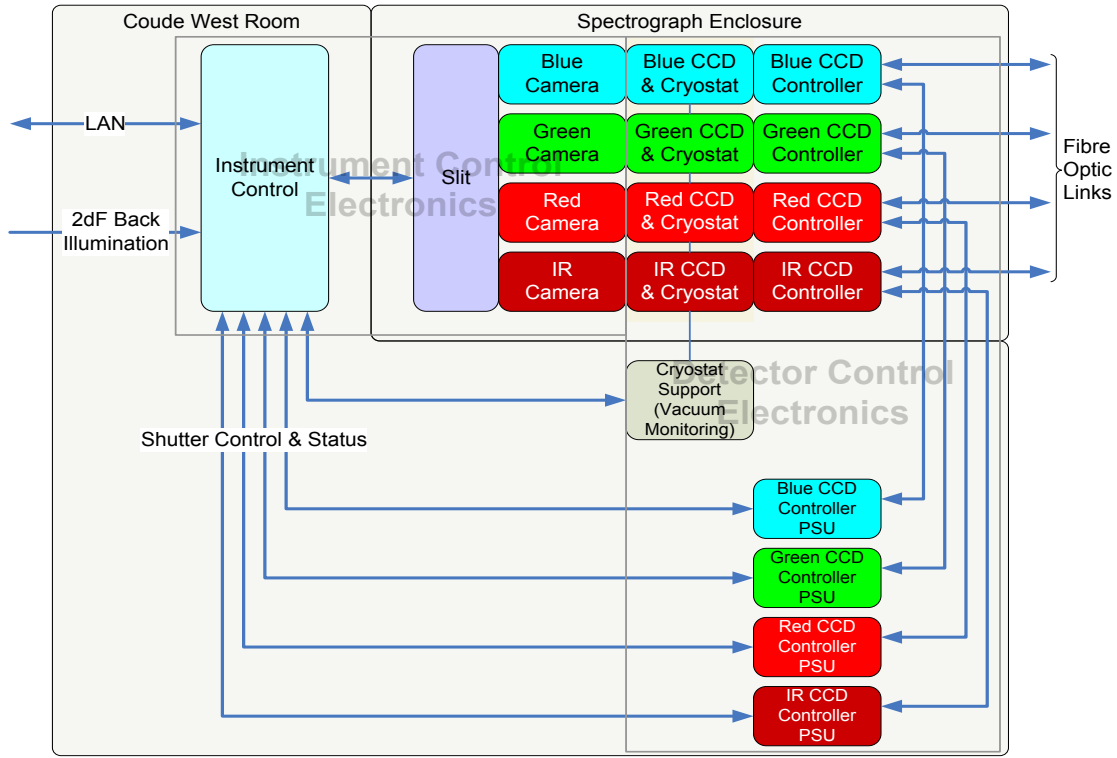

*Figure 19. Schematic overview of the HERMES Electronics Systems*

A CAN/CANopen system including a digital servo drive and digital and analog input and output was prototyped during the final design phase of the project. This has been available to allow software and hardware development and testing to proceed from an early stage in the project.

## 8. SOFTWARE AND DATA REDUCTION

### 8.1 Architecture and layout

Extensive use was made of existing software deployed as part of the 2dF/AAOmega instrument. This software was written using the AAO's DRAMA API[7] for the deployment of the original 2dF Instrument configuration, in 1994, but has been modified over the years for new spectrographs, detector systems and configurations.

For development purposes, the software for HERMES was broken into five major areas. The Observing system is responsible for overall control of the entire observing run-time (including the 2dF robot and CCD operations) and provides the main GUI. The HERMES Spectrograph Control task is responsible for control of the spectrograph itself. The AAO2 CCD detector software[8], shown in the figure below as CCD System, is used to run the detector systems. The Fiber Configuration software, not shown, is responsible for allocating fibers to objects as part of proposal preparation. A separate Data Reduction package provides pipeline reduction to the point of producing calibrated spectra.

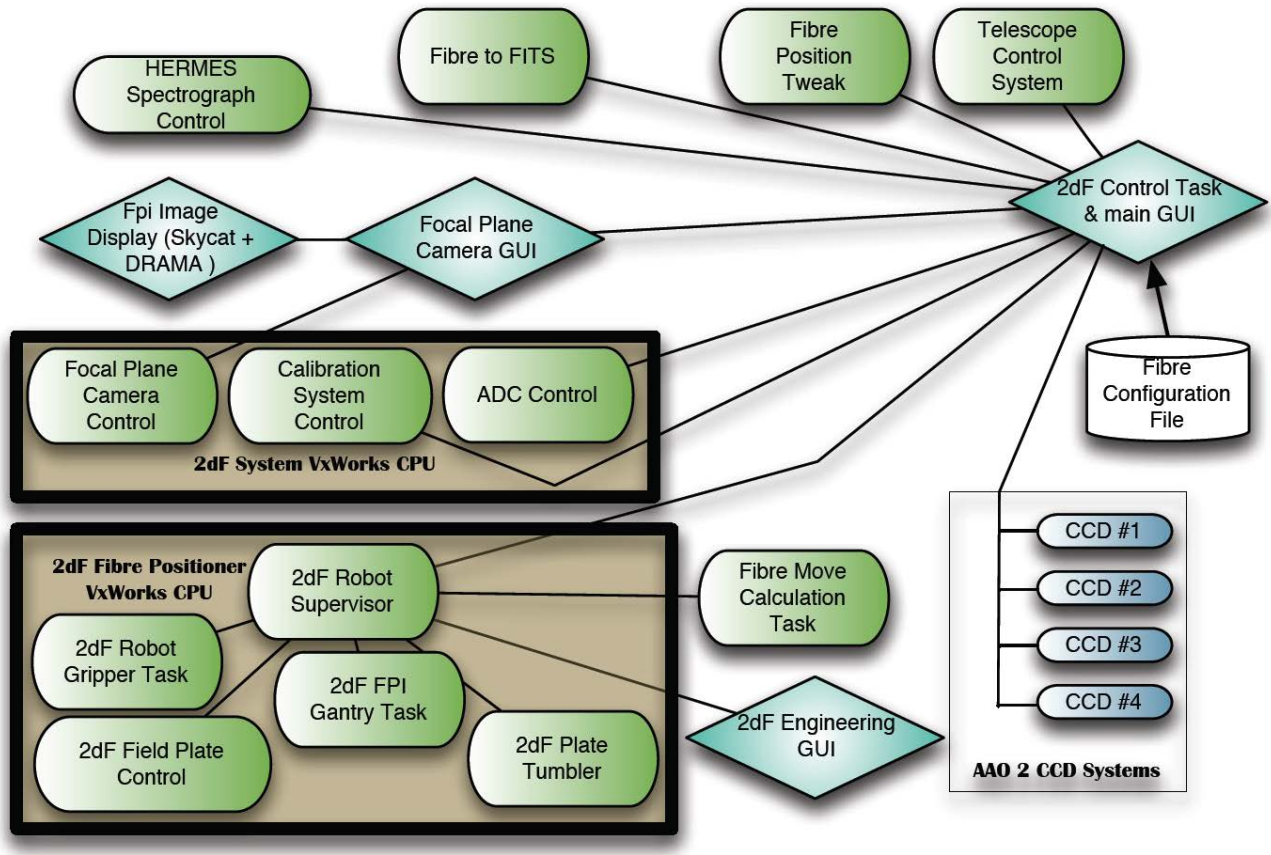

*Figure 20. Overview of the 2dF observing system software tasks when running in HERMES mode*

### 8.2 The HERMES Observing System

The existing 2dF Observing system has been modified to control HERMES in addition to AAOmega. The Observing system controls the 2dF Fiber Positioner, the Spectrographs, Detector Systems, 2dF Calibration system, 2dF Atmospheric Dispersion Corrector, the Telescope and a number of minor components. Each sub-system has its own DRAMA task. The AAO2 CCD detector software is itself an independent set of DRAMA tasks used by the 2dF software, as is the AAT's Telescope Control software. Overall control, software interlocking and synchronization are provided by the 2dF Control Task. This Observing system has proven flexible and adaptable since being originally commissioned in 1994, with the CCD System, Telescope System and Spectrographs having been changed during that time. For HERMES, we adapted the control task to implement a new mode of operation where the AAOmega Spectrograph DRAMA task is replaced by the new HERMES Spectrograph DRAMA task. Additionally, the control task and detector software were adapted to control four CCD systems, rather than the two previously supported. Switching between AAOmega mode and HERMES mode is via a command line option at system startup.

All sub-systems of the 2dF Observing system are required to implement simulation modes that allow for development of the Observing System without access to the instrument hardware. This was used extensively and as a result the 2dF Control Task was largely adapted to work with HERMES before the instrument itself was completed.

### 8.3 Spectrograph Control

The HERMES Spectrograph DRAMA task replaces the AAOmega spectrograph task when 2dF is running in HERMES mode. This task is responsible for communication with the spectrograph hardware and implementing high-level commands (DRAMA commands) for controlling the hardware. It provides instrument state for inclusion in FITS file headers and makes status information available via DRAMA parameters for inclusion in the system GUI.

An extensive hardware simulator has been provided using instrument simulator technology previously developed at the AAO[9]. This provides simulation of the CANBUS electronics hardware of the spectrograph at a level that requires no changes to the Spectrograph Task when running in simulation. Individual components of the instrument may be simulated at will. We used this extensively to advance development of the Spectrograph Task.

### 8.4 CCD Control

HERMES was the first AAO Instrument using 4 readouts (quadrants) per detector. It is also the first instrument using 4 detectors at once and using large detectors of 4kx4k pixels. Previous instruments were limited to 2 readouts per detector and 2 detectors at once. The AAO 2 CCD software was designed to support 4 readouts, but with no such detector configuration previously available, the implementation was not completed. We have now been able to complete this implementation, whilst at the same time optimizing the HERMES detector control modes. Adaption of the software to support 4 detectors running in one configuration was made early in the HERMES project, and we regularly operated four detectors in simulation whilst adapting the 2dF control task for HERMES. Due to hardware limitations, we only had 2 detector controllers available simultaneously before the HERMES instrument was moved to the AAT, so the final testing of the detector control configuration was not possible until then, but all high-risk areas had been tested and the final move to 4 detector controllers was completed without any issues.

### 8.5 Fiber Allocation

To observe a field on 2dF, you must allocate objects to fibers. This is done using a program known as "AAO Configure". Implemented originally for 2dF in 1994, it has been adapted to support the 6dF Instrument on the UK Schmidt Telescope and FLAMES on the VLT UT2, and to various changes in 2dF over the years. It has also been modified to implement a Simulated Annealing allocation method.

Minor changes were required for HERMES - basic instrument support and supporting the use of proper motions[10]. Proper motions had been supported in the FLAMES implementation, but had not been implemented with 2dF until now. The allocation algorithm was modified for HERMES, to optimize for fiber efficiency considerations such as cross talk

### 8.6 Data Reduction and Data Simulator

The AAO provides software to reduce the image data files produced by HERMES to the extent of generating calibrated spectra with (as far as practical) all instrument signatures removed. Science processing is left to the science teams, but in this case there was clear requirement to interface efficiently to the Galactic Archaeology abundance pipeline. The AAO's existing Fiber Instrument Reduction Package – 2dFdr, was adapted to support HERMES. This package had previously proven to be adaptable to different Instruments and such adaptations are now relatively easy to implement. The signal to noise requirements on HERMES for the Galactic Archaeology project, combined with the high resolution of the instrument, has meant fairly substantial changes to some sections of 2dFdr. In an effort to avoid the typical delay of 6 to 12 months after instrument commissioning before data reduction is optimal, we implemented a HERMES Data Simulator[11,12]; unfortunately, is was not able to capture a number of the characteristics of real HERMES data. The following significant improvements have been made to the software:

- Tramline tracking (tracing of the fiber paths on the detector) accuracy is improved to ~0.1pixel
- Optimal extraction routines have been extensively revisited, with more improvements underway.
- Routines are more stable (resulting in fewer pixels incorrectly marked bad)
- Variance estimation is much more accurate
- Tramline mapping in extremely sparse fields has been improved
- A b-spline scattered light option has been added to optimal extraction
- Improved Wavelength Calibration using an algorithm based on Whale Shark identification[15] and tank tracking to achieve wavelength calibration to ~0.1pixel
- Principle Component Analysis (PCA) Sky Subtraction[14]

Currently the Galactic Archaeology team is undertaking the analysis required to complete characterization of the data, to allow us to further optimize the reduction. 2dfdr development projects for the future include the use of twilight and dome flats to map the illumination of the lamp flats; the use of sky emission lines for determining throughput in each fibre, which is critical for accurate sky subtraction; development of a 2D deconvolution or "spectroperfectionism" algorithm for spectrum extraction; and removal of the VPH Littrow ghost.

## 9. SPECTROGRAPH PERFORMANCE

### 9.1 Spectrograph on-sky performance

HERMES on-sky commissioning was carried out over the months of October, November and December 2013. A range of observations were carried out and covered a lot of ground; determining readout speeds and combination of amplifiers, the required calibrations and their exposure times, PSF variations, fiber cross talk and scattered light levels, and overall instrumental resolution and throughput. This information allowed selecting the default instrument settings, to determine the magnitude limits, to set up the focus procedure and to confirm the appropriate alignment of the optics. A summary of the commissioning results is presented.

### 9.2 Spectral Resolution

HERMES provides two resolution modes, the nominal resolution mode of R ~ 28,000 and a higher resolution option utilizing a slit-mask (at the cost of 50% light loss). The HERMES nominal and mask mode spectral resolution was measured using the ThXe calibration lamp exposures.

### 9.3 Throughput

The throughput of the AAT+2dF+HERMES system was determined by measuring the detector counts over a range of standard star observations. The over all flux transformation equations (from star to detector) are given below for each channel:

Blue Channel:   e- per resolution element per hour = $10^{(-0.4(0.993*B - 24.05))}$
Green Channel:  e- per resolution element per hour = $10^{(-0.4(1.18*V - 26.25))}$
Red Channel:    e- per resolution element per hour = $10^{(-0.4(1.07*R - 24.98))}$
IR Channel:     e- per resolution element per hour = $10^{(-0.4(0.89*I - 22.33))}$

Note that in the above transformations, B, V, R, I refer to the Johnson B,V, R and I filters which are the nearest photometric bands to the four HERMES channels, and is based on the median seeing (1.5") conditions at Siding Spring Observatory. Table 1 gives the corresponding magnitude limit to achieve $10^4$ electrons (signal-to-noise = 100) per resolution element in one hour of exposure time. The total system efficiency meets the designed 10% throughput specification.

| Channel | Magnitude |
|---------|-----------|
| Blue    | B = 14.2  |
| Green   | V = 13.8  |
| Red     | R = 14.0  |
| IR      | IR = 13.8 |

Table 1. The magnitude limits for each HERMES channel to achieve $10^4$ electrons per resolution element, in the nominal HERMES resolution mode.

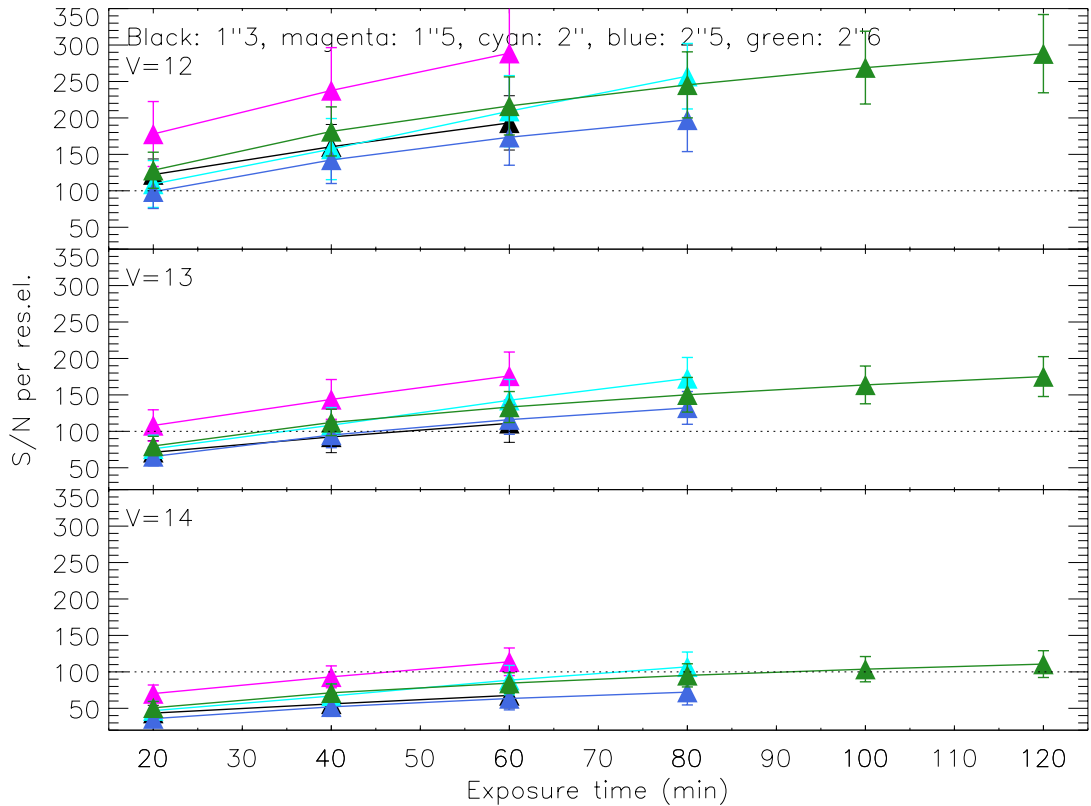

*Figure 22 shows that HERMES meets the design requirement that signal to noise per resolution element (S/N) should be 100 at V=14 in 1 hour of integration time, though this is modulated by the seeing. We have chosen stars with V magnitudes near 12, 13 and 14 (top, middle and bottom panels) from fields for which our observers recorded seeing of 1"1 (black), 1"4 (purple), 1"5 (cyan), 1"8 (blue), 2"2 (green) and 2"5 (dark green). We measure a typical S/N between 93 and 120 for stars with V=14 in 60 minutes of observing time when the seeing is 2" or better. Our observers extend the integration time by 20 minutes when the seeing is worse than 2" and by 60 minutes when the seeing is worse than 2"5, and it can be seen in the figure that this is effective at bringing the S/N up to the required levels.*

The use of the high-resolution mask reduces the flux by a factor of two, raising the limiting magnitude by 0.7mag. Overall the HERMES throughput performance meets the required specification to obtain a signal-to-noise of 100 per resolution element for a 14[th] magnitude star within one-hour exposure time. During lunar bright time, the sky background limits observations of targets fainter than 14[th] magnitude. In darker skies, HERMES could be used to observe targets down to 17[th] magnitude over extended exposure times to achieve S/N ratio between about 30-50.

### 9.4 Ghost and Scattered light

Scattered light in the form of dispersed background light from surface reflections is minimal in HERMES. The inter-slitlet regions on the detector can be used to measure the minimum signal and model the shape of the scattered light with a low order polynomial and subtract as part of the standard data reduction process.

A Littrow ghost, typical for VPH gratings used close to its Littrow configuration, affects HERMES data in all channels. The ghost images appear as small localized spots and fall spatially mirrored on the detector in a predictable manner forming the shape of the curved slit. The intensity of the ghost is at most 0.03% of the total flux in the originating fiber.

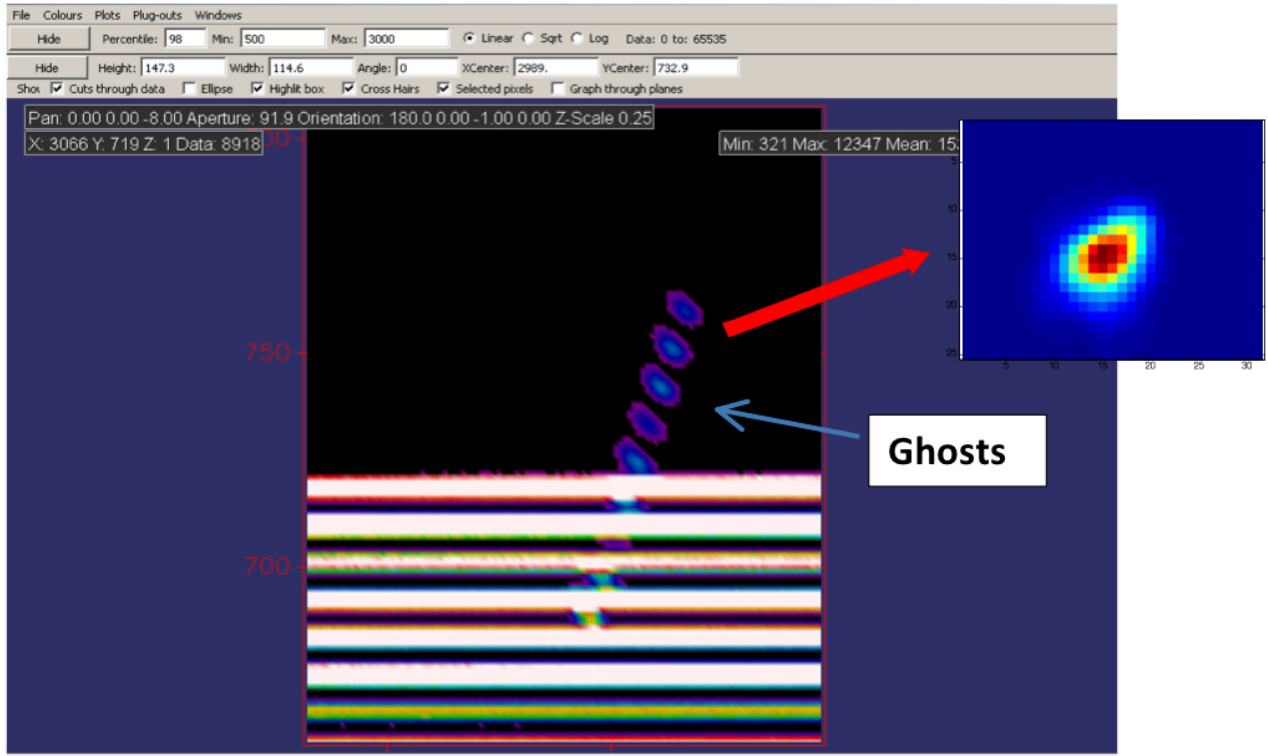

*Figure 23. A test flat field exposure with some of the fibers masked to identify the ghost.*

## 9.5 PSF variations and fiber cross talk
The optical PSF varies across the detector both spatially and spectrally in all channels. At the spatial center of the detector, the PSF convolved with a fiber image is circular with a typical spectral FWHM of 3.5pix at the blue wavelength end of the detector increasing to ellipse-shaped PSF convolved with a fiber images with a typical FWHM of 5.0pix at the red wavelength end of the detector. At the spatial top and bottom regions of the detector the elliptical PSF convolved with a fiber image is inclined in opposite directions. As shown in figures 24 and 25, this pattern matches the optical design of the instrument.

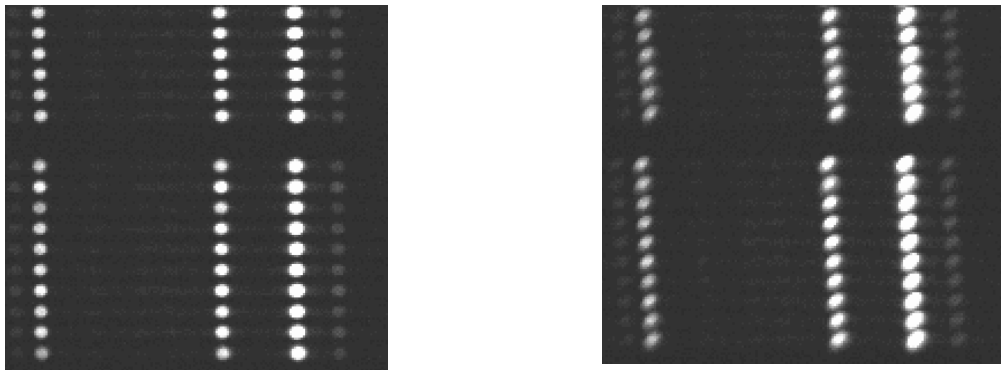

*Figure 24. (Left) is an image of an arc exposure zoomed in the center of the detector in the HERMES Red channel. The PSF convolved with a fiber images are round and the PSF convolved with a fiber images for adjacent fibers don't overlap significantly. (Right) is an image of the same arc exposure zoomed in the top of the detector. The PSF convolved with a fiber image is now an inclined ellipse and the PSF convolved with a fiber images for adjacent fibers do partly overlap.*

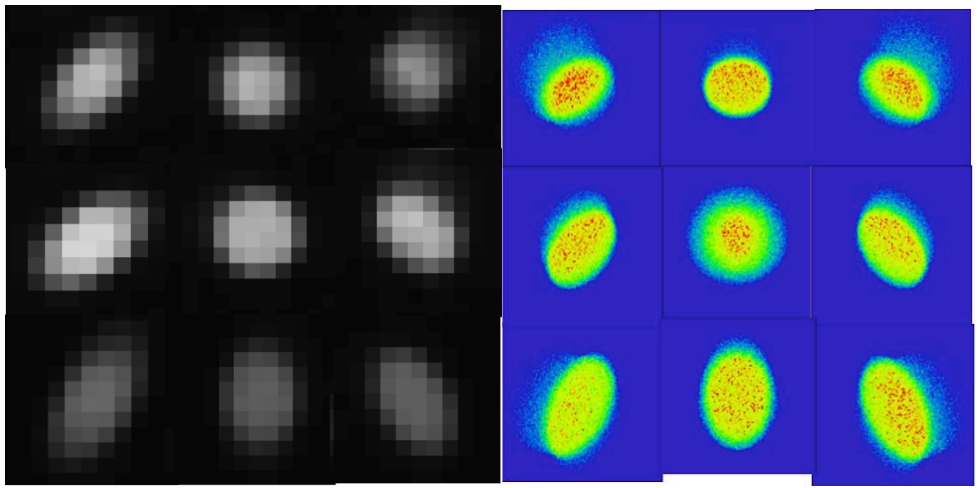

*Figure 25. (Left) shows the measured fiber images for the blue camera at the center and edges of the field. (Right) shows the predicted (from Zemax) fiber images at the center and edges of the field.*

The extension of the PSF in the spectral direction reduces the effective spectral resolution, but the most significant impact of this is on the fiber-to-fiber cross talk within each slitlet. The cross-talk level is such that 3-5% of flux from a fiber in the first and last 100 fibers affects its adjacent fibers. Given the limited hardware options to remove or reduce the observed cross talk, this level of crosstalk will require care during data reduction. Options being investigated include extracting the spectra using a 2D deconvolution and hard-edge clipping the affected data, which would cause a loss of signal but would ensure minimal contamination.

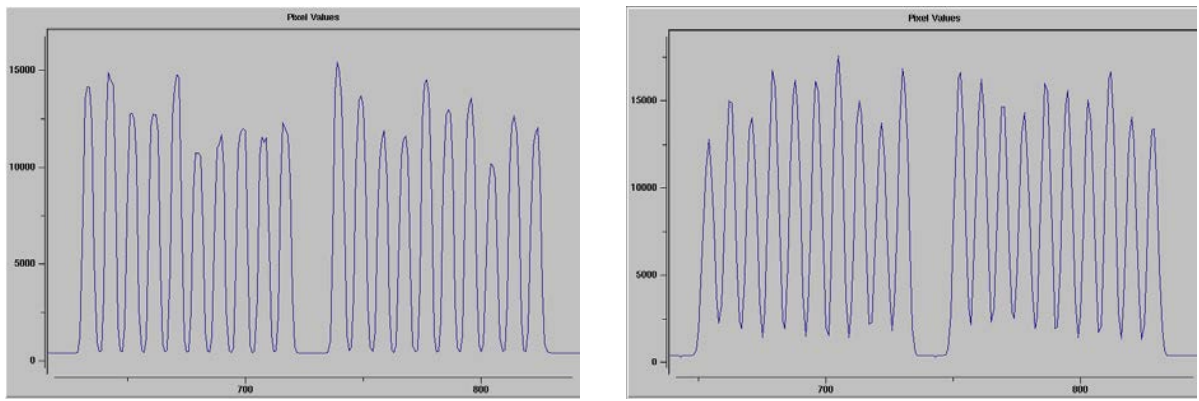

*Figure 26. (Left) plots flux vs. pixel position for a vertical cut across a fiber flat field image, zoomed in the center of the detector in the HERMES Red channel. Note the flux going down to the bias level between the two slitlets, and the flux between the individual fibers goes down similarly. (Right) plots the vertical cut in the same fiber flat, zoomed in the top of the detector. Note here the flux between the two slitlets go down to bias level, but flux level between individual fibers does not, indicating there is illumination from the adjacent fibers.*

## 9.6 HERMES Solar spectra
A short 30 sec exposure of the daytime sky was observed with 2dF+HERMES. The resulting data were reduced and extracted. The data was processed thru the GALAH analysis pipeline and was found to be in good agreement with solar spectra from other high-resolution spectrographs VLT+UVES and Keck+HIRES.

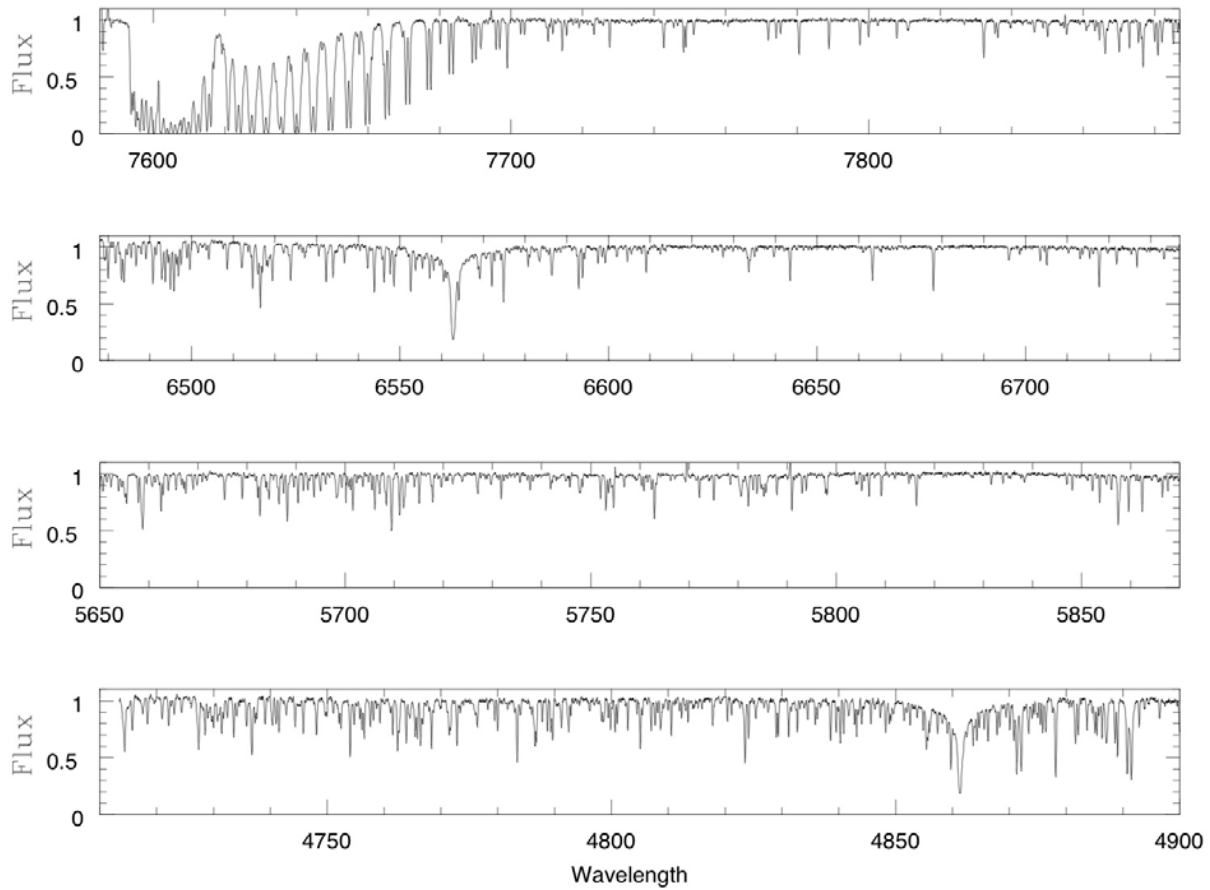

*Figure 27. shows the extracted and normalized solar spectra from the four HERMES channels*

## 10. GALAH Survey update
The primary science driver for the HERMES spectrograph is to carry out a large-scale Galactic Archaeology survey of the Milky Way[21]. The goal of Galactic Archaeology is to reconstruct the lost substructures of the proto-galaxy, thereby obtaining a detailed physical picture of the formation and evolution of the Milky Way. The GALAH (Galactic Archaeology with HERMES) survey is designed to carry out this task using the chemical tagging technique as a means to exploit the vast chemical inventory of one million disk stars[22]. The GALAH survey is an Australian-led project with members from 9 Australian institutes and several key international partners.

## 10.1 Pilot Survey

The GALAH team conducted a Pilot Survey from November 2013 to January 2014, to test survey strategy as well as carry out extended tests that were not possible during formal HERMES commissioning. These data address questions such as quality of flat-field correction from the various lamp options, repeatability of targets between fibers and plates, acquisition and guiding dependence on input catalogue parameters and the overall stability of calibrations. GALAH observations have contributed a total of 56 hours to acquire HERMES test data.

The Pilot Survey obtained spectra of ~13,000 individual science targets, including fundamental calibration stars, calibration open and globular star cluster fields and specific targeted science fields. These are used for testing the data reduction and analysis pipelines over the range of stellar parameters covered in the GALAH survey.

## 10.2 Observation summary

The GALAH survey was granted 75 nights of observing time over the period of February 2014 to January 2015 to begin its main observing campaign. In the first 56 nights of the main survey, GALAH observed 69884 stars in 192 main survey fields, as well as the previously mentioned HERMES test data. The GALAH Analysis Pipeline (GAP) conducts the required bookkeeping of the observations, data reduction checks, radial velocity determination, continuum normalization, and operation of the spectrum synthesis determination of abundances. A series of papers outlining the survey goals, observational strategies and detailed pipeline processing are in preparation, expected to be online by late 2014.

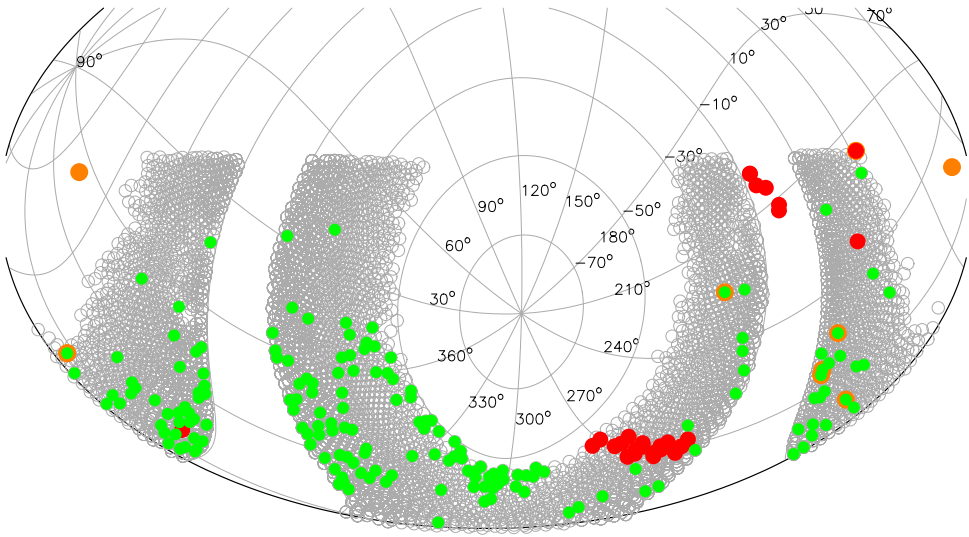

Figure 28. *shows the distribution across the sky of pilot survey (red), test (orange), main survey (green) and K2 (blue) fields. Main survey fields that also serve as test data are shown as green circles with an orange ring. There are 4303 field centers, shown as grey open circles, which are tiled into 6546 "configurations" containing 400 unique stars. The ultimate goal of a million-star sample requires observing 3200 of these configurations.*

## 11. CONCLUSIONS

The AAO has finished another very successful spectrograph, the HERMES spectrograph. The spectrograph is facilitating the GALAH Galactic Archeology survey, which is well on its way to achieve the goal of measuring abundances of 1,000,000 stars within the Milky Way. The instrument is performing to specification and the observations are underway.